\documentclass[
    ,final            % use final for the camera ready runs
  ]
  {aipproc}

\pdfoutput=1
\layoutstyle{8x11double}
\usepackage{amsmath}
\usepackage{amssymb}
\usepackage{xspace}
\usepackage{multirow}

\begin{document}

\title{Atmospheric Results from Super-Kamiokande}

%
% flux
% neutrons
% tau appearance 
% Three-flavor analysis + external constraints 
% Sterile oscillations
% Lorentz invariance 
% WIMPs from the galactic center
% Wimps from the Sun
%%


\author{Roger Wendell, for the Super-Kamiokande collaboration}{
  address={Kamioka Observatory, Institute for Cosmic Ray Research, University of Tokyo, Kamioka, Gifu, 506-1205, Japan} 
}

\begin{abstract}
Recent results from a 282 kiloton-year exposure of the Super-Kamiokande detector to 
atmospheric neutrinos are presented. 
The data when fit both by themselves and in conjunction with constraints from the 
T2K and reactor neutrino experiments show a weak, though insignificant, preference 
for the normal mass hierarchy at the level of$\sim$~1$\sigma$.
Searches for evidence of oscillations into a sterile neutrino have resulted in limits 
on the parameters governing their mixing, $ |U_{\mu 4}|^{2} <0.041 $ and $|U_{\tau 4}|^{2} < 0.18 $ 
at 90\% C.L.
A similar search for an indication of Lorentz-invariance violating oscillations 
has yielded limits three to seven orders of magnitude more stringent than existing measurements.
Additionally, analyses searching for an excess of 
neutrinos in the atmospheric data produced from the annihilation of dark matter particles in the galaxy and sun
have placed tight limits on the cross sections governing their annihilation and scattering.
\end{abstract}

\maketitle

%%%%%%%%%%%%%%%%%%%%%%%%%%%%%%%%%%%%%%%%%%%%
%% MAINMATTER
%%%%%%%%%%%%%%%%%%%%%%%%%%%%%%%%%%%%%%%%%%%%
%%\section{Introduction}

%%\include{full}

%% without the help of the BibTeX program. This could be used instead
%% of the above.
%%%%%%%%%%%%%%%%%%%%%%%%%%%%%%%%%%%%%%%%%%%

%With the recent discovery of non-zero $\theta_{13}$~\cite{Abe:2011sj,Abe:2011fz,Ahn:2012nd,An:2012eh} the analysis
%of atmospheric neutrino data becomes increasingly focused
%on addressing the remaining unknowns in the standard
%neutrino oscillation paradigm.
%Though the dominant sensitivity of the atmospheric neutrino
%sample is to $\nu_{\mu} \rightarrow \nu_{\tau}$ oscillations,
%the presence of matter along the path of upward-going neutrinos
%affects the $\nu_{\mu} \rightarrow \nu_{e}$ oscillation probability
%and enables the study of these open questions.
%Indeed, resonant enhancement of the oscillation probability
%is expected for multi-GeV neutrinos traversing the core and mantle regions of the earth.
%This, in combination with both oscillations driven by the solar oscillation
%parameters at sub-GeV energies and interference effects between the two domains
%at intermediate energies, gives the atmospheric neutrino sample
%sensitivity to the neutrino mass hierarchy, the octant of
%$\theta_{23}$, and the value of $\delta_{cp}$.
%This work presents the most recent analysis of the Super-Kamiokande (SK, Super-K)
%atmospheric data and its constraints on these open issues.

Atmospheric neutrino data at Super-Kamiokande (Super-K, SK) due to their wide 
variation in both energy and pathlength are sensitive to a variety of physical 
processes. 
In particular they are powerful probes of both standard neutrino oscillations 
and mixing induced by exotic scenarios such as sterile neutrinos or Lorentz-invariance 
violation. 
At the same time they represent the most serious background to searches for 
nucleon decays and neutrinos produced in the self-annihilation of DM particles 
gravitationally bound within the Sun or Milky Way.
Recent results from Super-K for a variety of these topics are presented below.
For a review of results in the modeling and use of atmospheric neutrinos 
the interested reader is referred to~\cite{Kajita:2004ga}.

\subsubsection{Detector and Data Set}

Super-Kamiokande uses 50 kilotons of ultra-pure water contained within a 41.4~m by 39.3~m cylindrical 
tank located at a depth of 2700 meters water equivalent in Gifu Prefecture, Japan for its measurements. 
The detector is optically separated into a target volume viewed by 11,146 inward-facing 20 inch
photomultiplier tubes (PMTs), which form the inner detector (ID), and an outer 
volume instrumented with 1,885 8 inch PMTs, that form the outer detector (OD). 
There is a 55~cm region between the ID and OD that is not instrumented.  
Cherenkov radiation produced by charged particles traveling through the detector's water 
is collected by the PMTs and used to reconstruct physics events.

\begin{figure}[htbp]
  \includegraphics[width=0.8\textwidth]{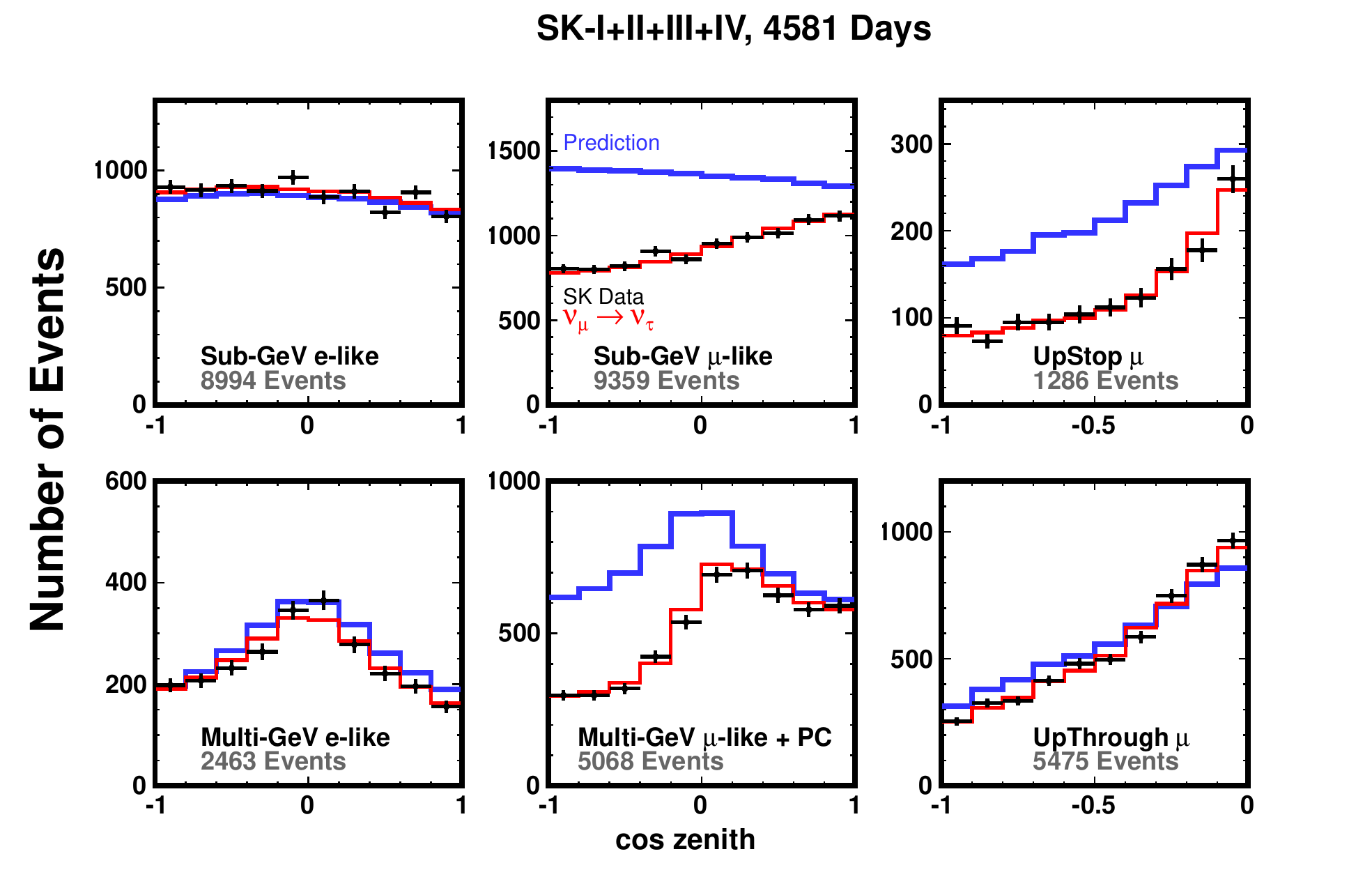}
  \caption{Super-K atmospheric neutrino data (points)and MC for a subset of the 19 samples used in the 
           oscillation analysis. 
           The MC prediction without neutrino oscillations is shown in blue 
           and the prediction including oscillations at the analysis best fit point 
           is shown in red. }
  \label{fig:zenith}
\end{figure}

Super-K has been in operation since 1996 and has collected 
neutrino data over four separate run periods in that interval. 
During the SK-I, SK-III, and SK-IV run periods the inner volume
photocathode coverage was 40\% while during the SK-II period, which was
instrumented with half the number of ID PMTs, it was 20\%.
After an accident in 2001 which destroyed half of the ID PMTs used 
during SK-I, the ID PMTs have been encased in a fiberglass 
reinforced plastic shell to prevent the production of shockwaves
in the event of a PMT implosion. 
In 2008 the detector electronics were upgraded to allow all PMT hits to 
be recorded, resulting in operations with near zero deadtime.
This upgrade marked the start of SK-IV, the current SK run period.
Details of the detector construction and calibration are presented in~\cite{Abe:2013gga}.

Atmospheric neutrino data are separated into three topologically distinct categories, 
fully contained (FC), partially contained (PC), and upward-going muons (Up$\mu$).
Fully contained interactions are those in which the primary event vertex 
has been reconstructed within the 22.5 kiloton ID fiducial volume, defined as the 
volume offset from the PMT wall by two meters, and deposit little energy in the OD.
These interactions are subdivided into analysis samples based upon the 
number of reconstructed Cherenkov rings (single- or multi-ring), the event energy (sub-GeV or multi-GeV), 
 and the particle ID (PID) of their most energetic ring ($e$-like or $\mu$-like).
Further subdivisions designed to enhance the charged current (CC) interaction content of 
certain samples, or to separate neutrino from antineutrino interactions are also performed.
Events with fiducial vertices but exiting particles causing PMT hits in the OD are 
classified as PC.
Based upon the light deposition in the OD, PC events are subdivided into ``stopping'' 
and ``through-going'' topologies to indicate whether or not the ID-exiting particle 
stopped within the OD. 
Upward-going muon events are produced by neutrino interactions in the rock outside of 
Super-K energetic enough to enter the detector from below and deposit energy in 
both the ID and OD. 
Events that traverse the entire ID volume are classified as ``through-going'' 
while those that do not are labeled ``stopping.'' 
The through-going sample is further separated into a ``showering'' and ``non-showering''
component based upon whether muon induced an electromagnetic shower within the ID. 
After all selections there are 19 analysis samples, some of which are shown 
in Figure~\ref{fig:zenith}.

Since the start of SK-IV it has become possible to search for the 2.2 MeV gamma ray 
following the capture of neutrons on protons, p(n,$\gamma$)d, in the detector water.
During this period 500$\mu$s of PMT data following a physics trigger are stored to 
enable offline searches for these events. 
However, because the photons are very low energy and produce a small number of 
PMT hits, extracting the neutron capture 
signal from background events induced by radioactive contaminants 
is done using a neural network procedure. 
It should be noted though that it is not possible to 
completely simulate all sources of potential background 
at these energies, so the background model is constructed using detector data sampled using a 
random trigger scheme with no hit or charge requirements. 

\begin{figure}[htbp]
  \includegraphics[width=0.35\textwidth]{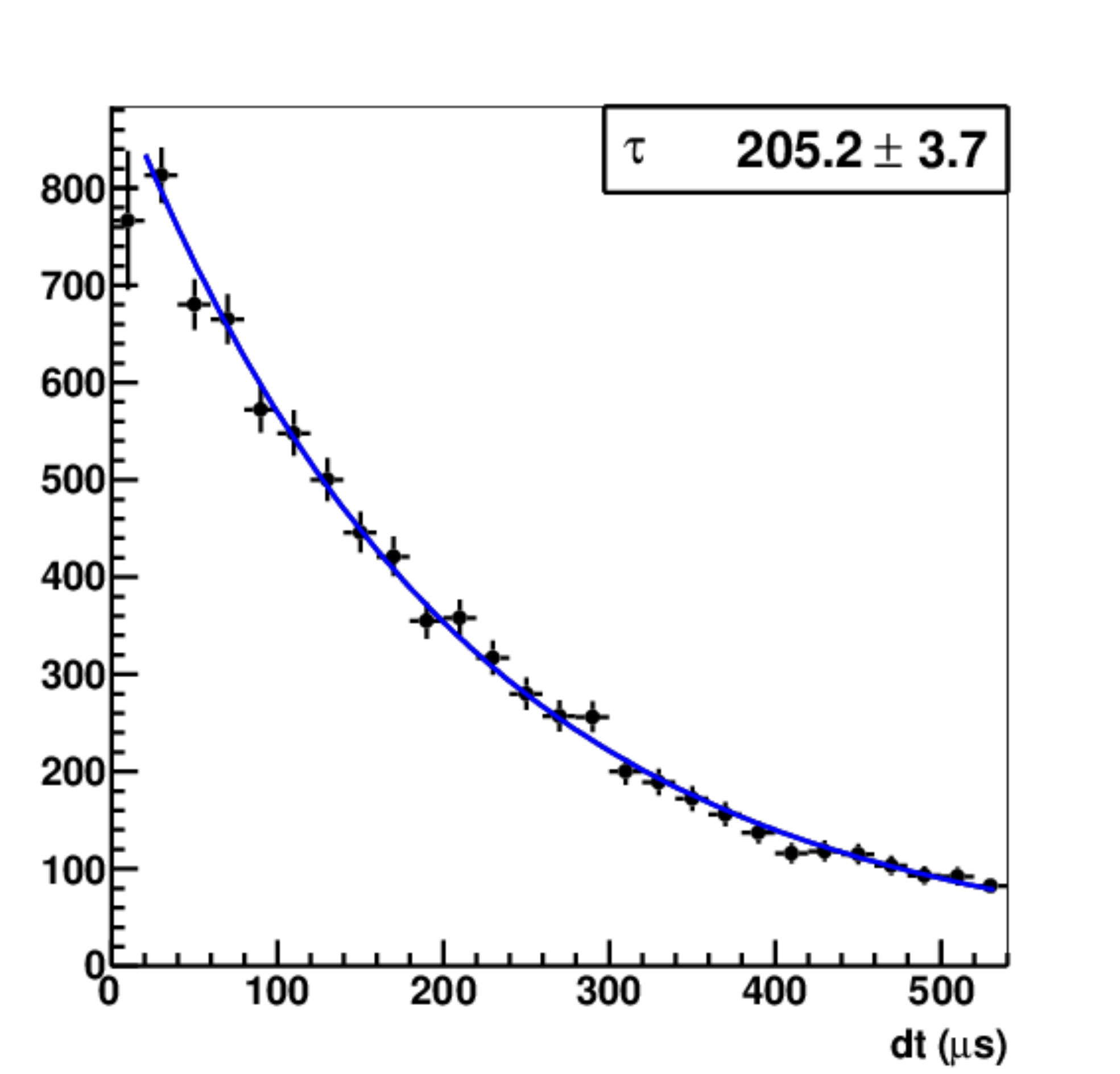}
  \caption{Time difference between tagged neutron capture events 
           and their associated primary event. 
           The blue line shows the fitted decay curve, which 
           yields a capture lifetime of 205.2$\mu$s.
           }
  \label{fig:neutron_dt}
\end{figure}

The neural network is built from 16 variables that isolate differences in the 
time, charge, and spatial distributions of hits originating from neutron capture events
and those from backgrounds. 
Since most neutrons are expected to capture within 200~cm of the primary 
neutrino interaction vertex, which is required to be at least this distance from 
from the PMT wall, variables that correlate the vertex position with the PMT activity 
are useful discriminants; Events from radioactive backgrounds in the detector materials 
may occur anywhere and often produce an isolated cluster of neighboring PMT hits, whereas 
neutron capture hits tend to be closer to the primary vertex with a spatial distribution 
that is consistent on average with the $41^{\circ}$ Cherenkov opening angle expected 
from low energy electrons. 
This neutron tagging procedure has an estimated efficiency of $20.5\%$ 
and incurs a misidentification rate of 0.018 neutrons per primary interaction. 
Figure~\ref{fig:neutron_dt} shows the time difference between selected neutron 
candidates and their primary interaction. 
Fitting this distribution yields a neutron capture time of $205.2\pm 3.7\mu$s,
in good agreement with existing measurements~\cite{Cokinos:1977zz}.
Application of this tagging to the atmospheric neutrino data set is expected 
to help distinguish CC antineutrino interactions, which are accompanied by 
a neutron at the primary vertex, from neutrino interactions and to help 
reduce backgrounds to proton decay searches.

\begin{figure}[htbp]
  \includegraphics[width=0.25\textwidth]{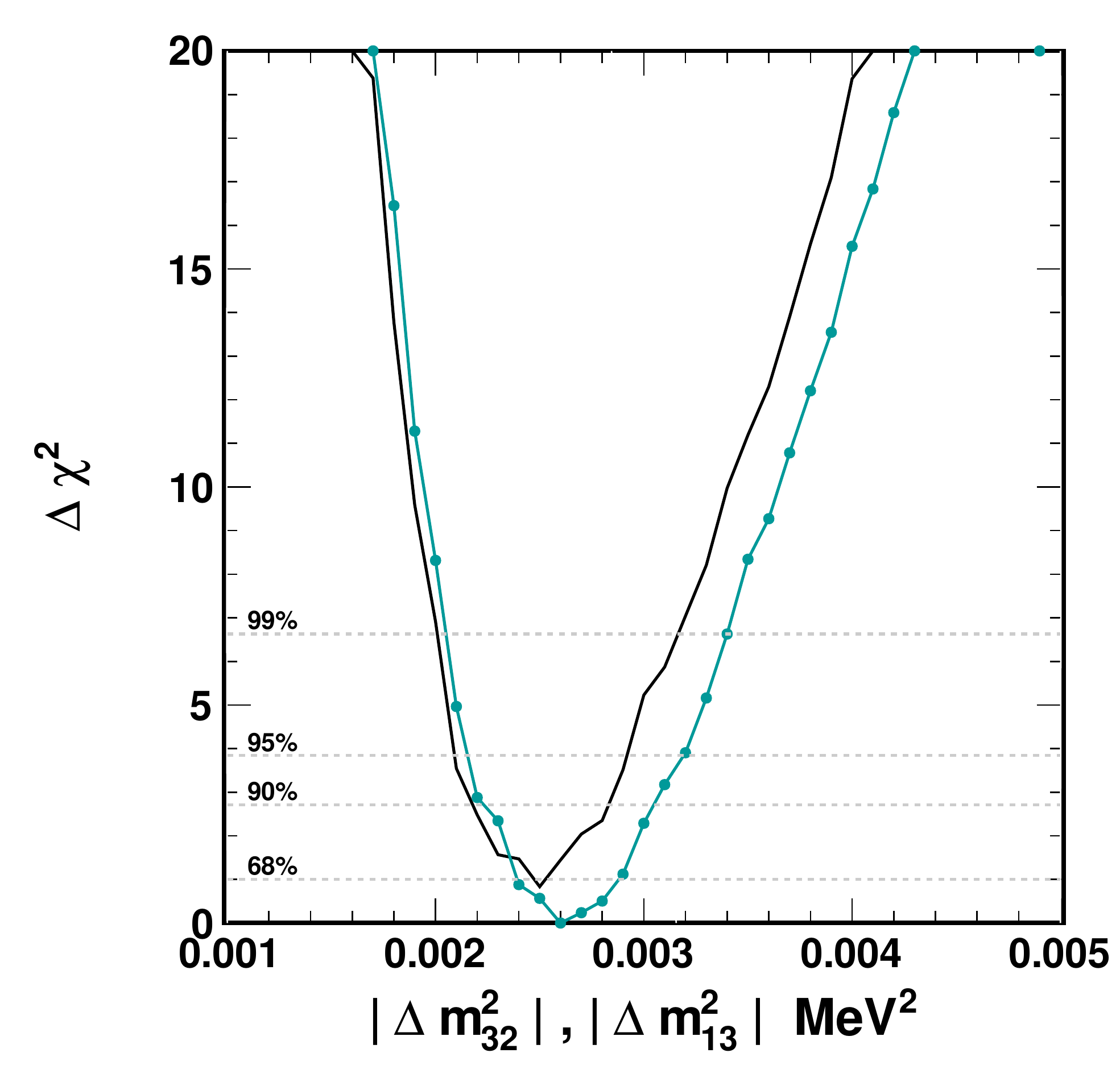}
  \includegraphics[width=0.25\textwidth]{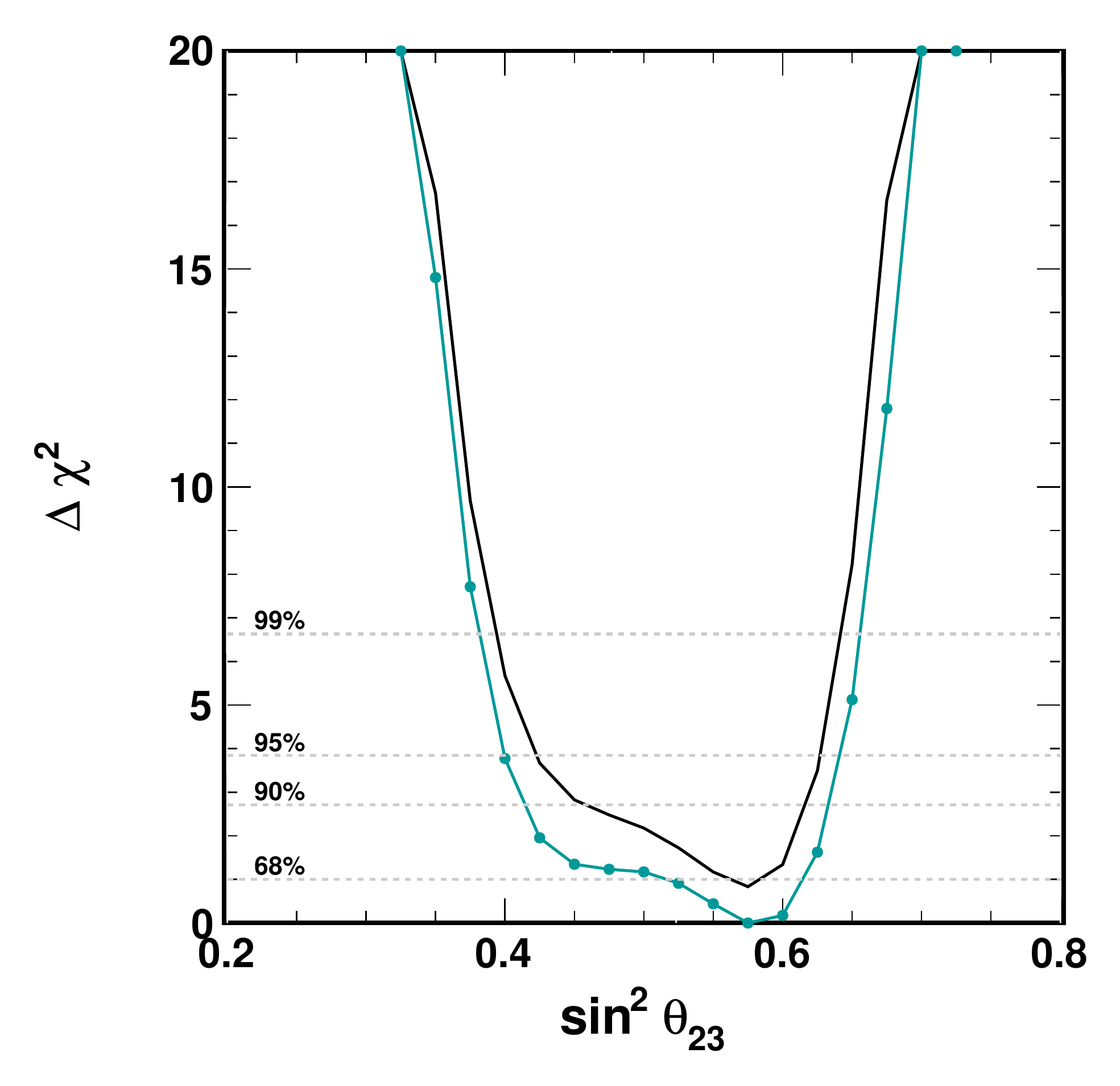}
  \includegraphics[width=0.25\textwidth]{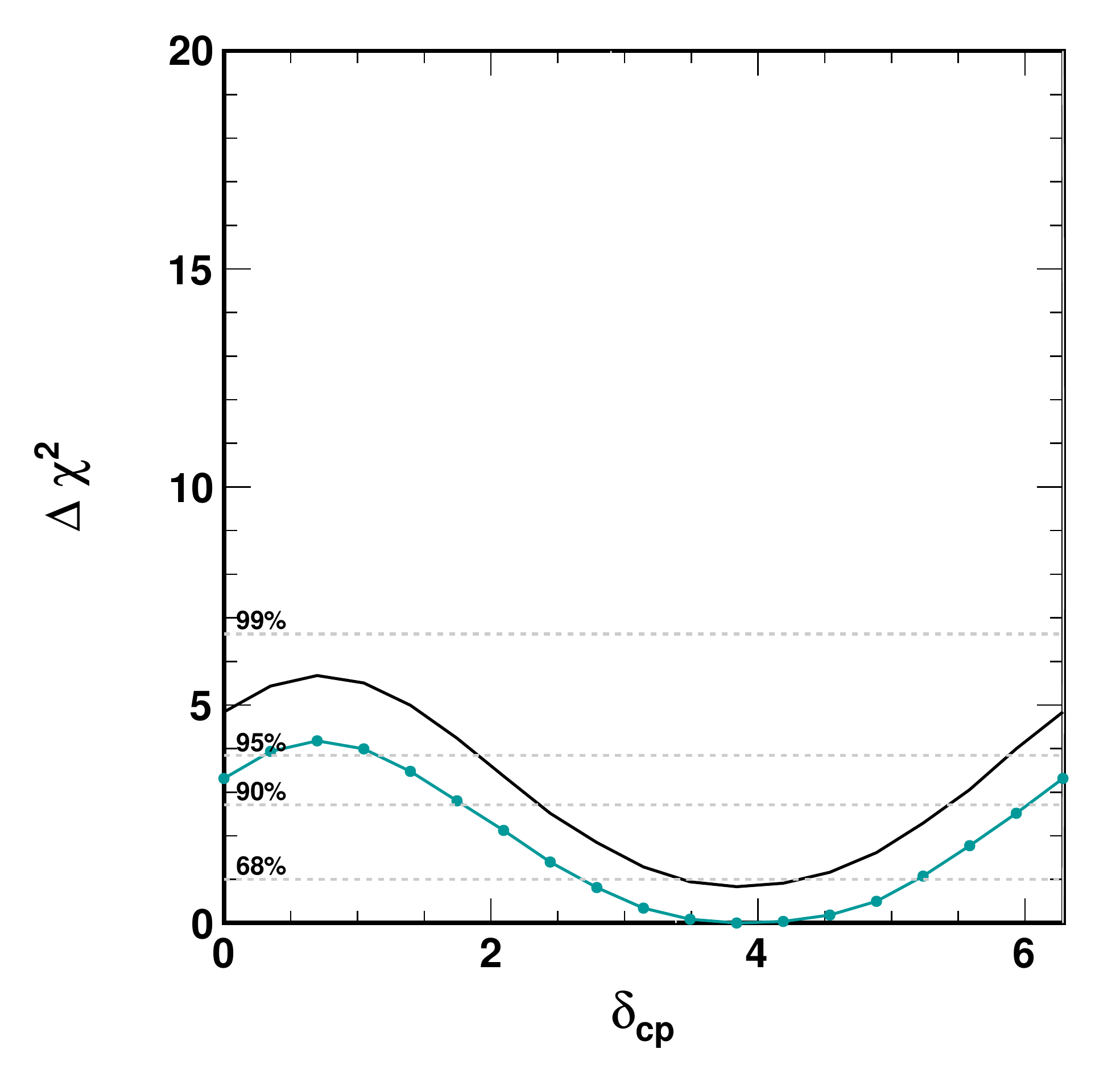}
  \caption{ Constraints on neutrino oscillation parameters from 
            the Super-K atmospheric neutrino data.
            Black lines denote the inverted hierarchy result, which 
            has been offset from the normal hierarchy result, shown in 
            blue, by the difference in their minimum $\chi^{2}$ values.}
  \label{fig:sk_cont}
\end{figure}

\begin{figure}
  \includegraphics[width=0.25\textwidth]{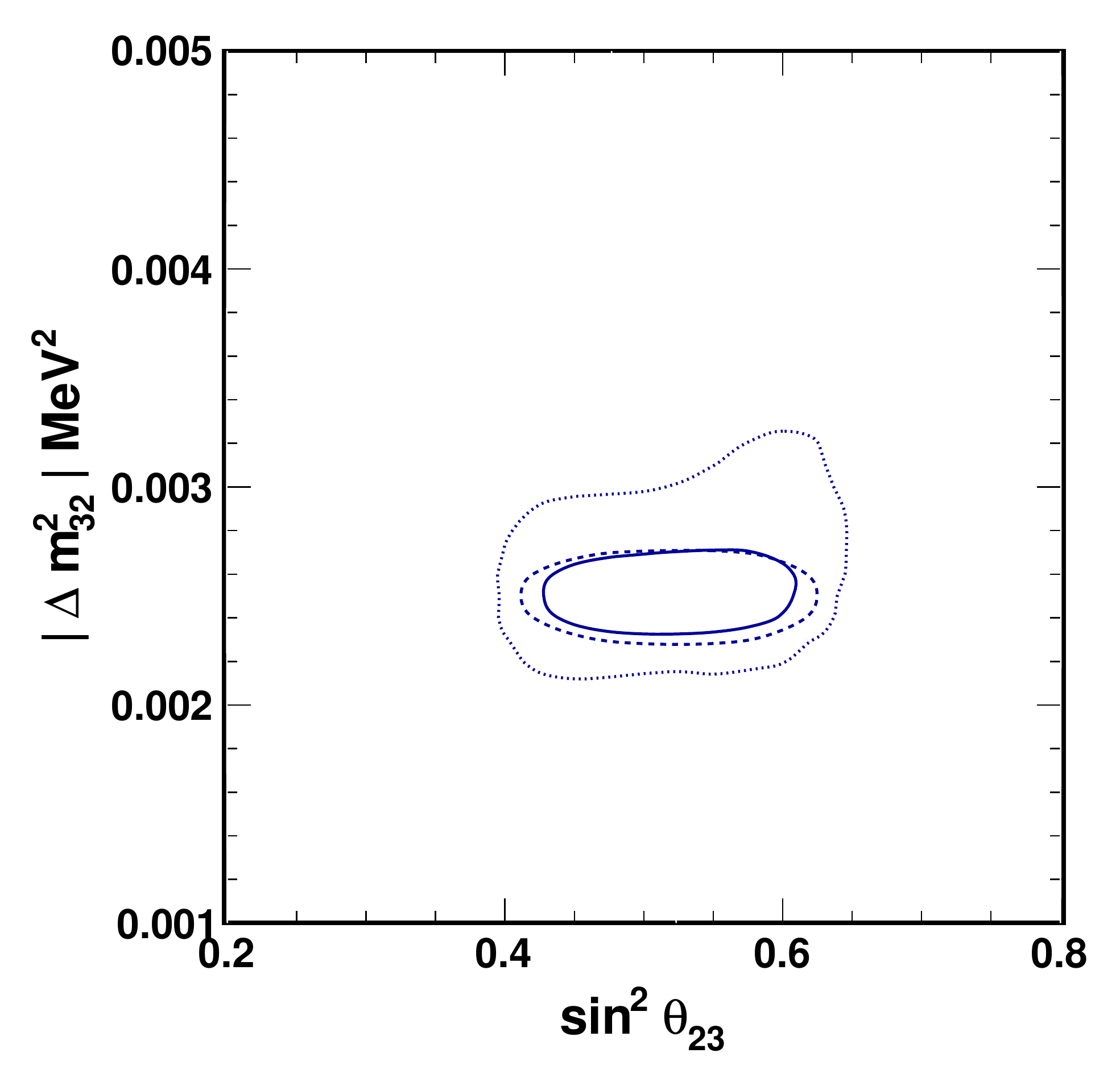}
  \includegraphics[width=0.25\textwidth]{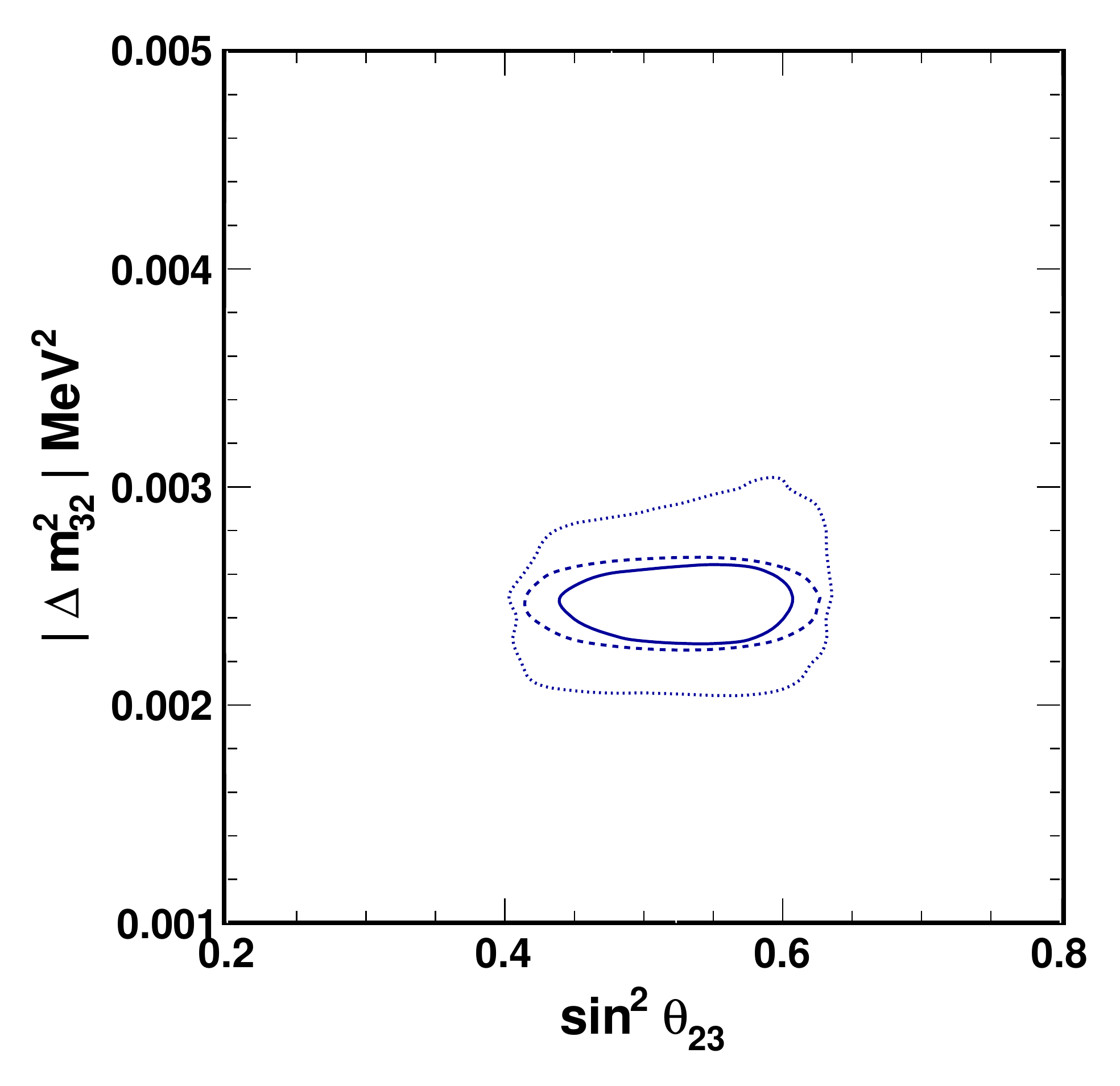}
  \includegraphics[width=0.25\textwidth]{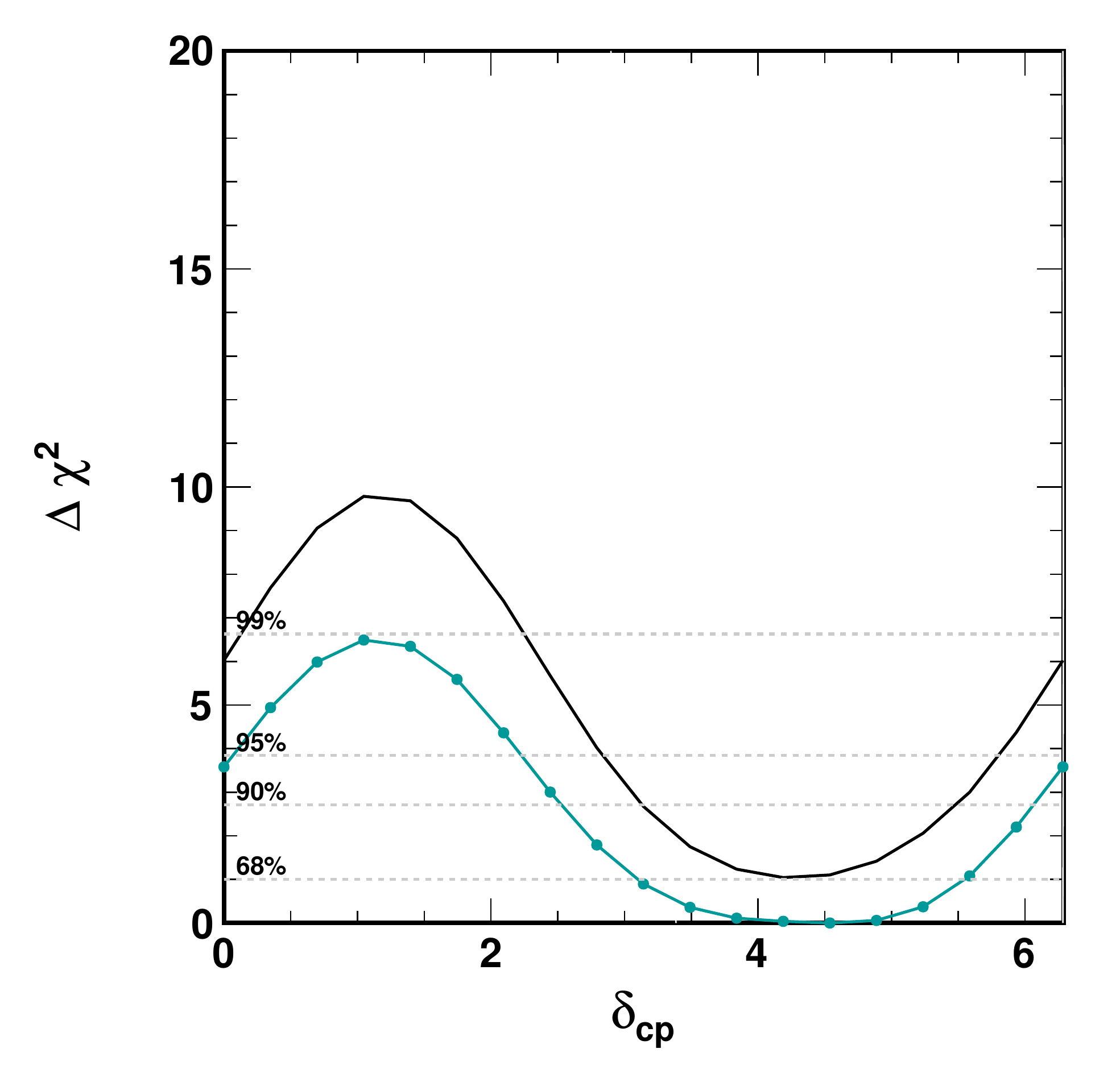}
  \caption{ Constraints on neutrino oscillation contours from a combined 
            fit of Super-K atmospheric neutrino data and a model of the 
            T2K experiment. The left two figures show two dimensional 
            constraints from the atmospheric neutrino data alone (dashed),
            the T2K model (dotted), and their combination (solid) for the 
            normal and inverted hierarchy, respectively. 
            The right figure shows the constraint on the $\delta_{CP}$ parameter 
            from the combined fit with the inverted hierarchy fit (black) 
            offset from the normal hierarchy fit (blue, markers) by the difference 
            in their minimum $\chi^{2}$ values.
           }
  \label{fig:skt2k_cont}
\end{figure}

\subsubsection{Standard Oscillations}

In the standard PMNS picture, atmospheric neutrino oscillations are dominated 
by the effects of the ``atmospheric'' mixing angle, $\theta_{23}$, and 
mass splitting $\Delta m^{2}_{32,13}$.
Broadly speaking these parameters drive oscillations of $\nu_{\mu}$ into 
$\nu_{\tau}$ and are readily seen as an absence of upward-going $\mu$-like events 
in the Super-K data~\cite{Abe:2011ph}.
However, observing the appearance of one neutrino flavor among a source 
dominated by another provides the cleanest evidence for oscillations
and has been the goal of several recent measurements~\cite{Abe:2013hdq,Agafonova:2013dtp,Adamson:2014vgd}.
Though in principal the detection of oscillated $\nu_{\tau}$ events 
at Super-K is straightforward since there are no $\nu_{\tau}$ in the 
primary atmospheric neutrino flux at the energies where oscillations are 
prominent, the 3.5 GeV production threshold of the $\tau$ lepton coupled with its 
prompt decay leads to high energy interactions with many light-producing particles 
that often resemble the deep inelastic scattering (DIS) interactions of the other neutrinos. 
For this reason extracting the $\nu_{\tau}$ signal requires some care to disentangle 
it from the backgrounds.

In the Super-K analysis~\cite{Abe:2012jj} a neural network (NN) is used to select 
CC $\nu_{\tau}$ interactions in which the outgoing lepton has decayed hadronically.
Though $\nu_{\tau}$ events are expected only in the upward-going data, the downward-going 
data provide an important constraint on the background, which composes 96\% of the 
events with a $\tau$-like NN classification.  
As a result, a two-dimensional PDF constructed from the complete zenith angle
and neural network output distributions is used to extract the $\tau$ signal. 
Using data from the first three Super-K run periods the analysis found 
$180.1\pm44.3\mbox{(stat)}^{+17.8}_{-15.2}\mbox{(syst)}$ $\nu_{\tau}$ interactions~\cite{Abe:2012jj}. 
This result is inconsistent with the hypothesis of no $\nu_{\tau}$ appearance 
at the $3.8\sigma$ level.

Despite being dominated by $\nu_{\mu} \leftrightarrow \nu_{\tau}$ oscillations, 
oscillations with the third neutrino, $\nu_{e}$, are expected to play a more 
important role due to the effects of $\theta_{13}$ and the Earth's matter. 
Indeed, between two and 10 GeV, matter induces a resonant enhancement of the 
$\nu_{\mu} \rightarrow \nu_{e}$ oscillation probability for upward-going neutrinos
traversing the core and mantle of the planet. 
Importantly, this enhancement depends both on the sign of the mass hierarchy 
and the CP state of the neutrino: for a normal (inverted) mass hierarchy the resonance 
exists for neutrinos (antineutrinos) only.
Since there are both neutrinos and antineutrinos in the atmospheric flux the ability 
to separate the two provides enhanced hierarchy sensitivity.   
Even without it, the resonance effect leads to a $\sim 12\% (5\%)$ increase in the 
rate of both single- and multi-ring multi-GeV upward-going $e$-like events 
assuming a normal (inverted) mass hierarchy. 
  
At these energies the increase in DIS cross section leads to complicated multi-ring 
event topologies.
Often the PID of the primary lepton is obscured by rings from charged hadrons, 
making it difficult to distinguish muons from electrons.  
In order to purify the multi-ring sample in CC $\nu_{e} + \bar\nu_{e}$ interactions, 
events whose most energetic ring's PID is $e$-like are subjected to a
likelihood selection which removes CC $\nu_{\mu} + \bar \nu_{\mu}$ and neutral current (NC)
interactions~\cite{Wendell:2010md}. 
This selection provides a passing sample that is $\sim 75\%$ pure in CC $\nu_{e} + \bar\nu_{e}$
interactions. 
It should be noted though that in order to combat the effects of low statistics and to 
provide additional systematic error constraints on the NC and $\nu_{\tau}$ components of 
the analysis samples, the rejected events are now used for the first time in the analysis. 
This additional sample is composed of roughly equal parts CC $\nu_{e}, \nu_{\mu}$, and NC 
interactions, with a $5\%$ contribution from $\nu_{\tau}$.

To improve sensitivity to the mass hierarchy, events in the passing sample are separated 
into $\nu_{e}$- and $\bar \nu_{e}$-like subsamples by a second likelihood procedure.
The likelihood is constructed to exploit the expected differences in both the number of 
expected charged pions and the amount of transverse momentum in the respective interactions 
(c.f.~\cite{maggie}). 
%(c.f.~\cite{Pik:2012qsy}). 
After selection the $\nu_{e}$($\bar\nu_{e}$)-like sample is $59.4\%$($21.0\%$) pure in 
CC $\nu_{e}$($\bar\nu_{e}$) interactions.   
A similar separation, based solely on the number of Michel electrons found, 
is performed for multi-GeV single-ring $e$-like events.
The respective purities after this cut are 62.8\% and 36.7\%.

Using these samples in conjunction with both $e$-like and $\mu$-like data 
from the remainder of the atmospheric data set, a simultaneous fit for 
$\theta_{23}$, $\Delta m^{2}_{23}$, and $\delta_{cp}$ is performed 
for each hierarchy assumption. 
During these fits the value of $\mbox{sin}^{2}\theta_{13}$ has been fixed to 
$0.025$ based on recent measurements~\cite{An:2012bu,Ahn:2012nd,Abe:2011fz} 
but its uncertainty is included as a systematic error. 
The ``1-2'' terms are treated in a similar fashion. 
A total of 154 systematic errors stemming from uncertainties in the 
flux and cross section models, as well as the detector's performance 
are included in the fit.  
Using the Super-K atmospheric neutrino data a weak preference for the 
mass hierarchy, $\chi^{2}_{IH} - \chi^{2}_{NH} = 0.9$, is found. 
Best fit information for this analysis is summarized in Table~\ref{tbl:atmfits}. 
One dimensional $\Delta \chi^{2}$ distributions for the oscillation variables 
are presented in Figure~\ref{fig:sk_cont}.
\begin{table}
\begin{tabular}{lcccc}
\hline \hline
Fit         & $\chi^{2}_{min}$ & $\delta_{cp}$ & $\mbox{sin}^{2}\theta_{23}$ & $\Delta m^{2}_{32,13}$ [eV$^{2}$] \\
\hline 
SK NH       &  559.8           &   3.84        &      0.57                   &  $2.6\times 10^{-3}$              \\ 
SK IH       &  560.7           &   3.84        &      0.57                   &  $2.5\times 10^{-3}$                           \\ 
\hline
SK$+$T2K NH &  578.2           &   4.19        &      0.55                   &  $2.5\times 10^{-3}$                           \\ 
SK$+$T2K IH &  579.4           &   4.19        &      0.55                   &  $2.5\times 10^{-3}$                           \\ 
\hline \hline
\end{tabular}
\caption{Summary of best fit information for fits assuming either the normal (NH) or inverted (IH) mass hierarchy using 
          either the Super-K atmospheric neutrino data only (517 d.o.f.), 
          or combining it with the T2K model (543 d.o.f.) described in the text.}
\label{tbl:atmfits}
\end{table}
Unfortunately, the magnitude of the resonance-enhanced oscillation probability 
described above carries a strong dependence on the value of $\mbox{sin}^{2}\theta_{23}$
and is similarly modulated, though to a lesser degree, by the value of $\delta_{cp}$.
If the former is 0.6 then the expected increase in the upward going multi-GeV event 
$e$-like rate is $17\%$ compared to $8\%$ if it is 0.4.
By using both $e$-like and $\mu$-like data Super-K is able to constrain 
the atmospheric mixing parameters, but since the direction of incoming atmospheric 
neutrino is unknown, its measurements are limited. 
Accordingly, the incorporation of external constraints on these parameters 
from more precise measurements is a means of increasing the sensitivity of the analysis.
%
%\begin{figure}[htbp]

Measurements of atmospheric mixing using accelerator neutrinos, such as those from 
the T2K~\cite{Abe:2013fuq} and MINOS~\cite{Adamson:2014vgd} experiments, provide 
tight constraints on both $\mbox{sin}^{2}\theta_{23}$ and $|\Delta m^{2}_{32}|$.
Since the T2K experiment uses Super-K as its far detector and the event simulation 
and reconstruction are common between beam and atmospheric neutrino events, it is
straightforward to build a model of the T2K experiment using only information 
in the literature and atmospheric analysis tools.
It is further possible to construct systematic error correlations between 
the atmospheric sample and a T2K model. 
Though MINOS~\cite{Adamson:2014vgd} provides important constraints on atmospheric 
mixing, notably $|\Delta m^{2}_{32}|$, the differences between it and Super-K make 
building an accurate model of the experiment more challenging. 
Below only a model of T2K is used to constrain the atmospheric neutrino analysis, 
but plans exist to extend those constraints to include MINOS and other experiments.

A model of the T2K experiment's $\nu_{e}$ appearance~\cite{Abe:2013hdq} and $\nu_{\mu}$ disappearance~\cite{Abe:2013fuq} 
analysis samples has been constructed using atmospheric neutrino tools. 
Using published beam flux information~\cite{Abe:2012av} atmospheric neutrino MC are reweighted 
to produce a beam MC. 
The beam MC is subjected to the same selection criterion outlined in the 
references to produce model $\nu_{e}$ and $\nu_{\mu}$ samples. 
Data are digitized from the references. 
T2K has published estimates of its systematic errors in terms of changes in the 
sample event rates and these have been recast into the Super-K style of 
systematic error treatment (c.f.~\cite{Wendell:2010md}).
Due to its low statistics and for ease of analysis only one analysis bin 
is used for the model $\nu_{e}$ sample; For the model $\nu_{\mu}$ sample the same binning as the 
reference is used. 

After construction of the systematic errors and analysis samples, independent fits of 
each model sample were performed to verify the model reproduced the same parameter constraints
as T2K's published results. 
These samples are then fit together with the 19 Super-K analysis samples, assuming 
full correlation between relevant systematic error parameters, such as those in the cross section model.
Contours from the combined fit are presented in Figure~\ref{fig:skt2k_cont} 
and the best fit information is listed in Table~\ref{tbl:atmfits}.
After inclusion of the T2K model, the preference for the normal mass hierarchy 
improved slightly to $\chi^{2}_{IH} - \chi^{2}_{NH} = 1.2$, but 
remains insignificant. 
Though a preference is seen in both the Super-K and combined analysis for $\delta_{cp}\sim 3\pi/2$
CP-conservation is allowed at at least the 90\% confidence level.

\newcommand{\dm}{\ensuremath{\Delta m^{2}}\xspace}
\newcommand{\dmsq}[1]{\ensuremath{\dm_{#1}}\xspace}
\newcommand{\dmnew}{\dm}
\renewcommand{\th}[1]{\ensuremath{\theta_{#1}}\xspace}
\newcommand{\sn}[1]{\ensuremath{ \sin^{2}(\theta_{#1}) }\xspace }
\newcommand{\snt}[1]{\ensuremath{ \sin^{2}(2\theta_{#1}) }\xspace }
\newcommand{\Hlv}{\ensuremath{H_{LV}}\xspace}

\newcommand{\at}{\ensuremath{a^{T}}\xspace}
\newcommand{\ctt}{\ensuremath{c^{TT}}\xspace}
\newcommand{\ats}[1]{\ensuremath{\at_{#1}}\xspace}
\newcommand{\ctts}[1]{\ensuremath{\ctt_{#1}}\xspace}
\newcommand{\as}[1]{\ensuremath{\left(a^{T}_{#1}\right)^*}\xspace}
\newcommand{\cs}[1]{\ensuremath{\left(c^{TT}_{#1}\right)^*}\xspace}

\newcommand{\val}[2]{\ensuremath{#1 \; \mathrm{#2}\xspace}}
\newcommand{\sci}[2]{\ensuremath{#1 \times 10^{#2}}\xspace}
\renewcommand{\Re}{\operatorname{Re}}
\renewcommand{\Im}{\operatorname{Im}}

\subsubsection{Exotic Mixing Scenarios}

\begin{figure}[htbp]
  \includegraphics[width=0.35\textwidth]{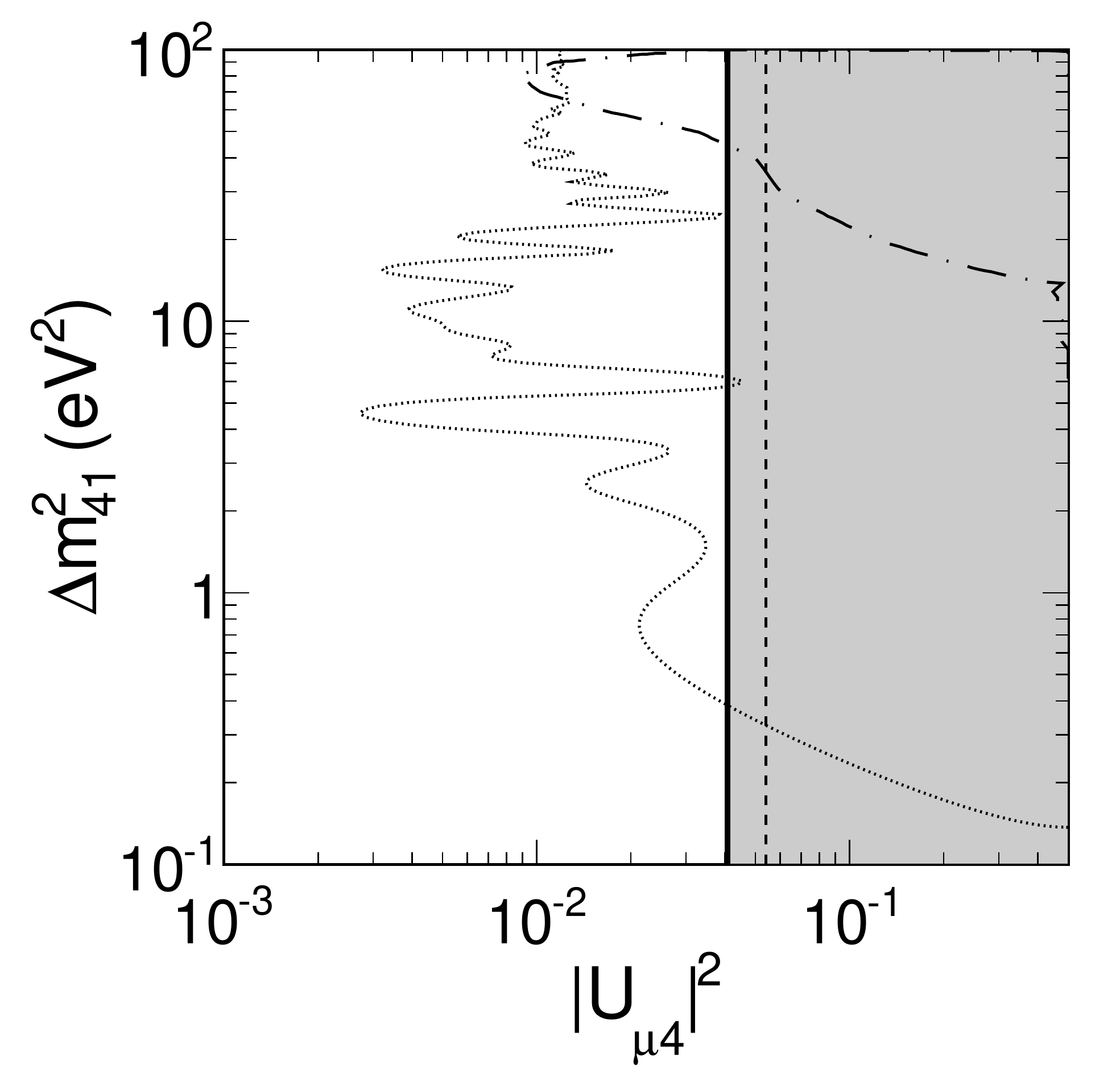}
  \includegraphics[width=0.35\textwidth]{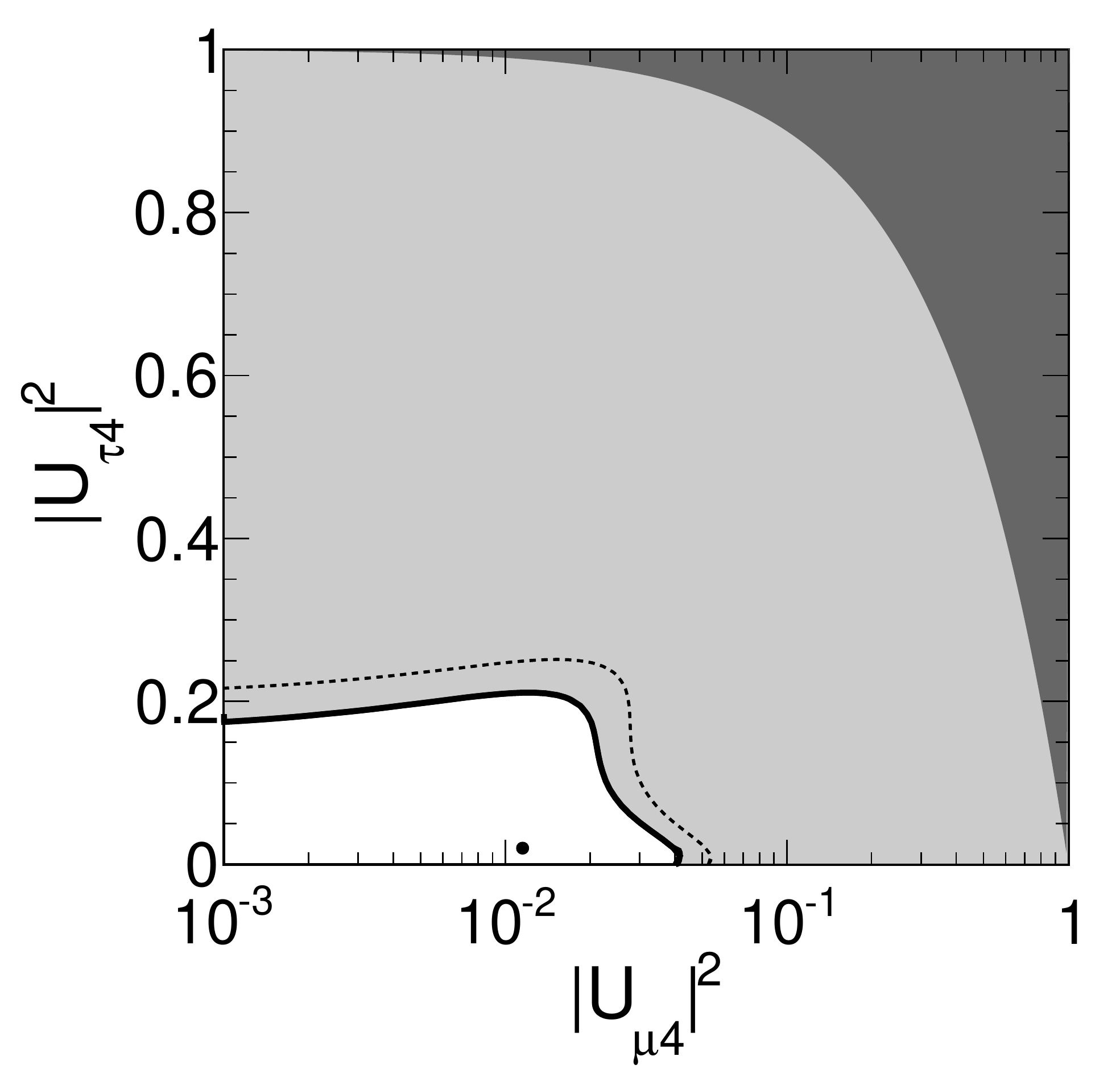}
  \caption{Constraints on sterile neutrino oscillations assuming
           the neutron density in the Earth is zero (left) and  
           $\nu_{e}$s do not participate in any oscillations (right).
           The solid (dashed) line indicates the 90\% (99\%) limit 
           with the exclusion region shaded in gray. 
           Limits at 90\% C.L. from a joint analysis of MiniBooNE and SciBooNE data~\cite{Cheng:2012yy}
           as well as from the CCFR experiment~\cite{Stockdale:1984cg} appear as the dotted 
           and dash-dot curves, respectively.}
  \label{fig:sterile}
\end{figure}

Though the standard paradigm of neutrino oscillations driven by two 
mass differences, $\Delta m^{2}_{12} \sim 10^{-5} \mbox{eV}^{2}$ 
and $\Delta m^{2}_{32} \sim 10^{-3} \mbox{eV}^{2}$, is well established,
there are a number of experimental hints for a third mass difference, 
$\Delta m^{2}_{s} \sim 1 \mbox{eV}^{2}$ (c.f.~\cite{Aguilar:2001ty,Aguilar-Arevalo:2013pmq,Huber:2011wv,Mention:2011rk})
that does not fit into this picture.
Based on constraints from measurements of the $Z^{0}$ decay width, it is known 
that the existence of a fourth (or higher) neutrino state to explain these hints 
implies that it does not participate in the weak interaction. It is is therefore 
termed ``sterile.'' 
If there are oscillations into such a state, its lack of interaction 
can induce reductions and spectral distortions in the atmospheric 
neutrino flux at Super-K.

As an interferometric effect atmospheric neutrino oscillations are
similarly sensitive to Lorentz-invariance violation (LV).
While there is currently no evidence for the violation of this symmetry, 
its fundamental role in both the standard model of particle physics 
and general relativity make it the target of critical experimental 
investigation.
Indeed, an observation of LV would provide access to Planck scale physics~\cite{Colladay:1996iz,Colladay:1998fq,Kostelecky:2003fs}, an energy scale far out of reach of collider experiments. 
In this section analyses searching for these types of exotic oscillation 
effects are presented.

\subsubsection{Sterile Neutrino Oscillations}

Atmospheric neutrinos can provide 
useful constraints on mixing into sterile states as they are essentially insensitive to the exact number of 
sterile neutrinos and for $\Delta m^{2}_{s} > 0.1 \mbox{eV}^{2}$ the oscillations 
are sufficiently fast that they are insensitive to the precise value of the mass splitting.
In general, the addition of sterile states to standard PMNS oscillations leads to 
a mixing matrix of the following form, 

\begin{equation}
U = \left(\begin{array}{ccccc}
U_{e1}    & U_{e2}    & U_{e3}    & U_{e4}    & \cdots \\
U_{\mu1}  & U_{\mu2}  & U_{\mu3}  & U_{\mu4}  & \cdots \\
U_{\tau1} & U_{\tau2} & U_{\tau3} & U_{\tau4} & \cdots \\
U_{s1}    & U_{s2}    & U_{s3}    & U_{s4}    & \cdots \\
\vdots    & \vdots    & \vdots    & \vdots    & \ddots \\
\end{array}
\right),
\end{equation}
\noindent where ellipses denote entries for models with more than one additional state. 
Further, since sterile neutrinos do not experience NC interactions they are 
subject to an additional effective potential,
\begin{equation}
V_{s}  = \pm  (G_F/ \sqrt{2}) \mbox{diag}(0, 0, 0, N_n, \ldots), 
\end{equation}
\noindent when traveling through matter which is not felt by the active neutrinos.
Here $N_n$ is the local neutron density and $G_F$ the Fermi constant.
To simplify the oscillation computations the number of sterile states is taken 
to be one and it is assumed there is no $\nu_{e} \leftrightarrow \nu_{s}$ mixing, 
i.e. $U_{e4}=0$. 
In this context oscillation probabilities for the analyses below have been 
calculated based on~\cite{Maltoni:2007zf}.

\begin{table}
\begin{tabular}{lcccc}
\hline \hline
Fit                & $|U_{\mu 4}|^{2}_{best}$ & $|U_{\tau 4}|^{2}_{best}$ & $|U_{\mu 4}|^{2}_{lim}$ & $|U_{\tau 4}|^{2}_{lim}$ \\ 
\hline 
No-$\nu_{e}$       &    0.012       &      0.021     &  --       &  $<0.18$       \\ 
Sterile$_{Vacuum}$ &    0.016       &      --        & $<0.041$  &   --           \\ 
\hline \hline
\end{tabular}
\caption{Summary of best fit information for sterile neutrino oscillations 
         for fits assuming either no $\nu_{e}$ oscillations 
         or lack of sterile matter effects (Sterile$_{Vacuum}$). 
         Best fits as well as limits at 90\% C.L. are shown. 
         The minimum $\chi^{2}$ for the No-$\nu_{e}$  and Sterile$_{Vacuum}$
         fits was 531.1 and 532.1 for 480 d.o.f., respectively.}
\label{tbl:sterilefits}
\end{table}

The analysis is divided into two pieces based on two approximations with 
different domains of applicability.
In the first analysis, it is assumed that there is no mixing between $\nu_{e}$ 
and the other active neutrinos, that is $\theta_{13} = \theta_{23} = 0$. 
Under this approximation atmospheric neutrinos are sensitive to 
both $|U_{\mu 4}|$ and $|U_{\tau 4}|$ at the expense of providing a 
slightly biased estimate of the former. 
Since $|U_{\mu 4}|$ decreases the overall $\nu_{\mu}$ survival probability 
and therefore produces a reduction in the normalization of the $\mu$-like samples,
their correlation with the $e$-like samples through systematic uncertainties in the flux
model
ultimately leads to a bias towards lower values of this parameter, due to the 
now missing $\theta_{13}$-induced $e$-like events discussed above. 
Constraints on $|U_{\tau 4}|$ are nonetheless possible as it introduces 
distortions in the shape of the higher energy $\nu_{\mu}$-rich PC 
and Up$\mu$ samples through the sterile matter effect.
The second analysis assumes regular PMNS mixing with a sterile neutrino 
but takes the neutron matter density to be zero. 
While this assumption removes the effects of matter on sterile neutrino oscillations, 
it allows for an unbiased measure of $|U_{\mu 4}|$. 

Both analyses use 18 of the 19 Super-K analysis samples for a total of 
480 degrees of freedom. 
Standard oscillation parameters are unvaried in the fit and have been 
fixed to 
$\Delta m^{2}_{32} = 2.51\pm 0.10 \times 10^{-3} \mbox{eV}^{2}$, 
$\mbox{sin}^{2}\theta_{23}  = 0.514 \pm 0.055$, 
$\mbox{sin}^{2}2\theta_{13} = 0.095 \pm 0.01$,  
$\Delta m^{2}_{21} = 7.46\pm 0.19 \times 10^{-5} \mbox{eV}^{2}$, 
$\mbox{sin}^{2}\theta_{12}  = 0.305 \pm 0.021$, with 
$\delta_{cp} = 0$ and assuming a normal mass hierarchy. 
The quoted uncertainties are treated as systematic errors during the
fit for the sterile mixing parameters. 
After performing fits to both of the oscillation scenarios 
no evidence for oscillations with $\Delta m^{2}_{s} \sim 1 \mbox{eV}^{2}$ 
nor for spectral distortions induced by sterile matter effects 
is found. 
Limits placed on the sterile mixing parameters as well as 
best fit information is summarized in Table~\ref{tbl:sterilefits}.
Allowed regions from the two analyses appear in Figure~\ref{fig:sterile}.

\subsubsection{Lorentz Violation in Neutrino Oscillations}

\begin{figure}[htbp]
  \includegraphics[width=0.30\textwidth]{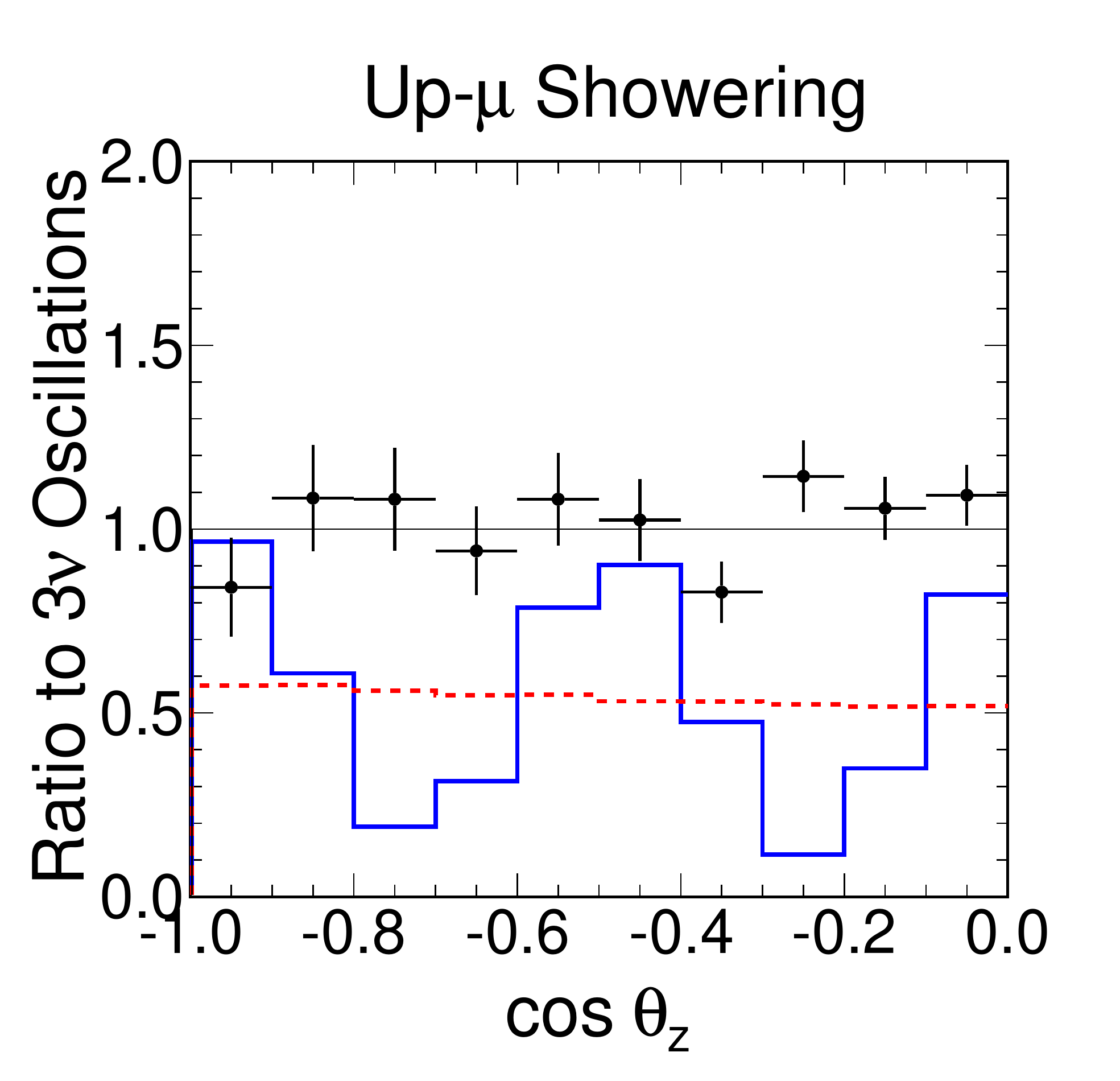}
  \caption{Illustration of the effects of LV on neutrino oscillations for the 
           the Up$\mu$ showering sample.  
           Data points are shown relative to the prediction from 
           standard oscillations as a function of zenith angle.
           The ratio of LV oscillations to standard oscillations 
           is shown in blue (red) for $ \ats{\mu\tau} = 10^{-22}$ GeV
           (  $ \ctts{\mu\tau} = 10^{-22}$ ).
           }
  \label{fig:liv}
\end{figure}

The effects of LV on atmospheric neutrinos may manifest as 
sidereal variations in their oscillations 
or as distortions in the oscillated spectra~\cite{Kostelecky:2011gq,Kostelecky:2003xn,Kostelecky:2003cr}.
Super-K has undergone a search for the latter within the framework 
of the standard model extension (SME)~\cite{Kostelecky:2003fs}, an observer independent 
effective field theory which contains the standard model, general relativity, 
and all possible LV operators. 
Within the SME the usual Hamiltonian describing neutrino propagation 
appears with a LV component,
\begin{align}
\Hlv =& 
\left(\begin{array}{ccc}
0          & \ats{e\mu}     & \ats{e\tau} \\
\as{e\mu}  & 0            & \ats{\mu\tau} \\
\as{e\tau} & \as{\mu\tau} & 0  \\
\end{array}\right) \nonumber\\
& -E 
\left(\begin{array}{ccc}
0          & \ctts{e\mu}     & \ctts{e\tau} \\
\cs{e\mu}  & 0            & \ctts{\mu\tau} \\
\cs{e\tau} & \cs{\mu\tau} & 0  \\
\end{array}\right),
\end{align}
\noindent where $\mbox{a}_{\alpha\beta}^{T}$ and  $\mbox{c}_{\alpha\beta}^{TT}$
are complex coefficients for isotropic LV operators. 
Generally the $\mbox{a}_{\alpha\beta}^{T}$ parameters produce oscillation effects 
proportional to the neutrino propagation distance, $L$, while 
the  $\mbox{c}_{\alpha\beta}^{TT}$ induce effects that depend on $LE$,
where $E$ is the neutrino energy. 
Given the form of these effects, the large variation in atmospheric neutrino 
pathlengths (from $\sim 10$ to $13,000$~km) and energy (from 100 MeV to 10 TeV and above) 
make them particularly sensitive probes.
However, it is for precisely this reason that perturbative approaches
such as~\cite{Diaz:2009qk} cannot be used and the full SME Hamiltonian must be 
diagonalized. 
Accordingly, oscillation probabilities for the analysis below have been obtained 
following~\cite{JSDiaz}.

The effects on neutrino oscillations from even modest amounts of LV can be dramatic and 
generally appear in the multi-GeV (both $e$-like and $\mu$-like), PC, and Up$\mu$ samples. 
As an example, Figure~\ref{fig:liv} shows the ratio of the zenith angle distribution of the 
Up$\mu$ showering data, whose events have both long flight lengths and the highest energy 
of the Super-K samples, relative to the expectation assuming standard oscillations.
A blue (red) line shows the effect of LV oscillations as the same ratio using the  
MC prediction assuming $ \ats{\mu\tau} = 10^{-22}$ GeV
(  $ \ctts{\mu\tau} = 10^{-22}$ ), both of which show large departures from standard 
oscillations. 
In the search for LV in the atmospheric neutrino data the standard oscillation 
parameters are held fixed at the same values used in the sterile oscillation search above,
with the exception of $\delta_{cp}$, which is allowed to vary freely in the fit. 
Each LV coefficient is then fit for both its real and imaginary components while 
holding all other LV parameters at zero.  
The results are listed in Table~\ref{tab:lvresults}.
No evidence for LV has been seen in any of the fits and has resulted both in limits 
three to seven orders of magnitude stronger than existing measurements~\cite{Katori:2012pe,Katori:2013jca}
and has also established new limits, in particular on the  $\ats{\mu\tau}$ and $ \ctts{\mu\tau} $ parameters.

\begin{table}
\renewcommand{\tabcolsep}{2pt}
\newcommand{\tsp}{\hspace{1.2em}}
\centering
\begin{tabular}{l@{\tsp}c@{\tsp}cc@{\tsp}cc@{\tsp}c@{\tsp}c}
\hline\hline
\multicolumn{2}{l}{LV Parameter}           &
\multicolumn{2}{c}{Best Fit}               &
\multicolumn{2}{c}{$95\%$ Upper Limit}     \\
\hline
\multirow{4}{*}{$e\mu    $} & $\Re\left(\at \right)$ & $1.0 \times 10^{-23} $ & GeV  & $1.8 \times 10^{-23} $ & GeV  \\
                            & $\Im\left(\at \right)$ & $4.6 \times 10^{-24} $ & GeV  & $1.8 \times 10^{-23} $ & GeV  \\
                            & $\Re\left(\ctt\right)$ & $1.0 \times 10^{-28} $ &      & $1.1 \times 10^{-26} $ &      \\
                            & $\Im\left(\ctt\right)$ & $1.0 \times 10^{-28} $ &      & $1.1 \times 10^{-26} $ &      \\
\hline                                                                                                               
\multirow{4}{*}{$e\tau   $} & $\Re\left(\at \right)$ & $2.2 \times 10^{-24} $ & GeV  & $4.1 \times 10^{-23} $ & GeV  \\
                            & $\Im\left(\at \right)$ & $1.0 \times 10^{-28} $ & GeV  & $2.8 \times 10^{-23} $ & GeV  \\
                            & $\Re\left(\ctt\right)$ & $1.0 \times 10^{-28} $ &      & $1.2 \times 10^{-24} $ &      \\
                            & $\Im\left(\ctt\right)$ & $4.6 \times 10^{-25} $ &      & $1.4 \times 10^{-24} $ &      \\
\hline                                                                                                               
\multirow{4}{*}{$\mu\tau $} & $\Re\left(\at \right)$ & $3.2 \times 10^{-24} $ & GeV  & $6.5 \times 10^{-24} $ & GeV  \\
                            & $\Im\left(\at \right)$ & $1.0 \times 10^{-28} $ & GeV  & $5.1 \times 10^{-24} $ & GeV  \\
                            & $\Re\left(\ctt\right)$ & $1.0 \times 10^{-28} $ &      & $5.8 \times 10^{-27} $ &      \\
                            & $\Im\left(\ctt\right)$ & $1.0 \times 10^{-27} $ &      & $5.6 \times 10^{-27} $ &      \\
\hline\hline
\end{tabular}
\caption{Summary of the results of the six fits for Lorentz-violating parameters where the real and imaginary parts of each parameter are fit simultaneously. The best fit parameters and their  95\% upper limits are shown. }
\label{tab:lvresults}
\end{table}

\begin{figure}[htbp]
  \includegraphics[width=0.45\textwidth]{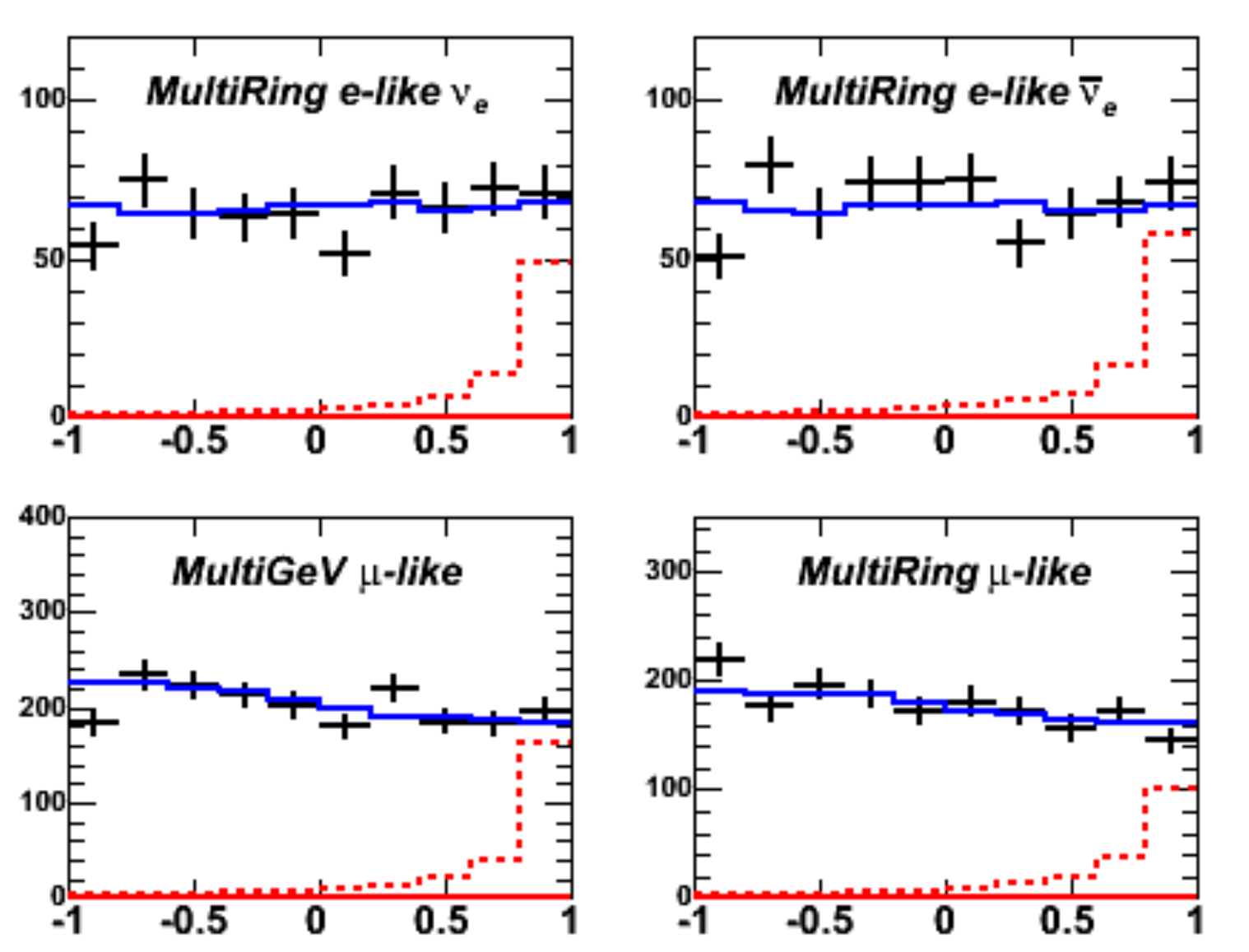}
  \caption{Illustration of the expected signal at Super-K from 5 GeV WIMP
           annihilation around the galactic center.
           The horizontal axis shows the cosine of the angle to the galactic center. 
           Data, atmospheric background MC, and the WIMP signal appear as the 
           markers, blue line, and red dashed line, respectively.
           Though the analysis 
           uses 18 analysis samples for the fit, only the four particularly relevant to 
           this channel are shown.
           The signal normalization has been exaggerated for visibility.
           }
  \label{fig:gcwimpsignal}
\end{figure}

\subsubsection{Indirect WIMP Searches}

While atmospheric neutrinos have been primarily discussed in the context 
of a signal source in the above, they also serve as a background to 
searchers for rare phenomena such as nucleon decays and dark matter (DM)
annihilation into neutrinos. 
Indeed, weakly interacting massive particles, WIMPS, are a promising 
candidate for cold dark matter as the weakness of their interactions 
provides a natural explanation of the relic abundance of DM~\cite{Steigman:1984ac,Jungman:1995df,Bertone:2004pz}.
If such particles exist it is possible that their annihilation or decays 
could produce standard model particles, such as neutrinos or particles whose 
decay chains include neutrinos.
In this scenario large gravitational potentials, such as the Milky Way or the Sun, 
can trap WIMPs, providing an arena for their collisions.
The existence of such a source would manifest in Super-K atmospheric neutrino 
data as a neutrino excess coming from a particular direction on the sky, with 
energies as large as the WIMP mass.
Searches for evidence of WIMP annihilation in the Sun and the galactic center (GC)
are described here.

\begin{figure}[htbp]
  \includegraphics[width=0.40\textwidth]{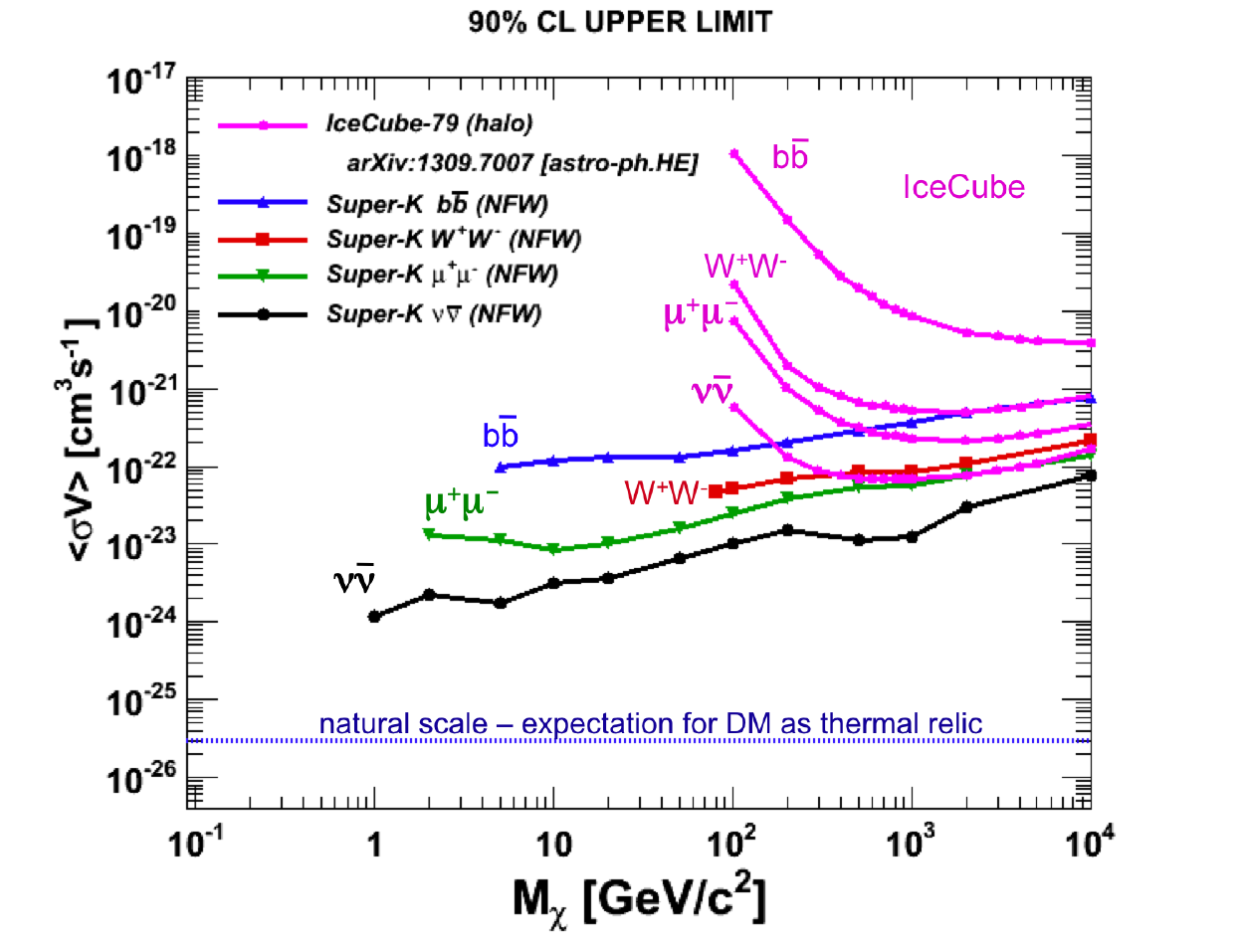}
  \caption{Upper limits on the WIMP self-annihilation cross section averaged over 
           the density distribution found in~\cite{Navarro:1995iw} for various assumed masses 
           and annihilation modes. Limits from the IceCube experiment~\cite{Aartsen:2013mla} are also shown in pink. }
  \label{fig:gcwimp}
\end{figure}

\begin{figure}[htbp]
  \includegraphics[width=0.40\textwidth]{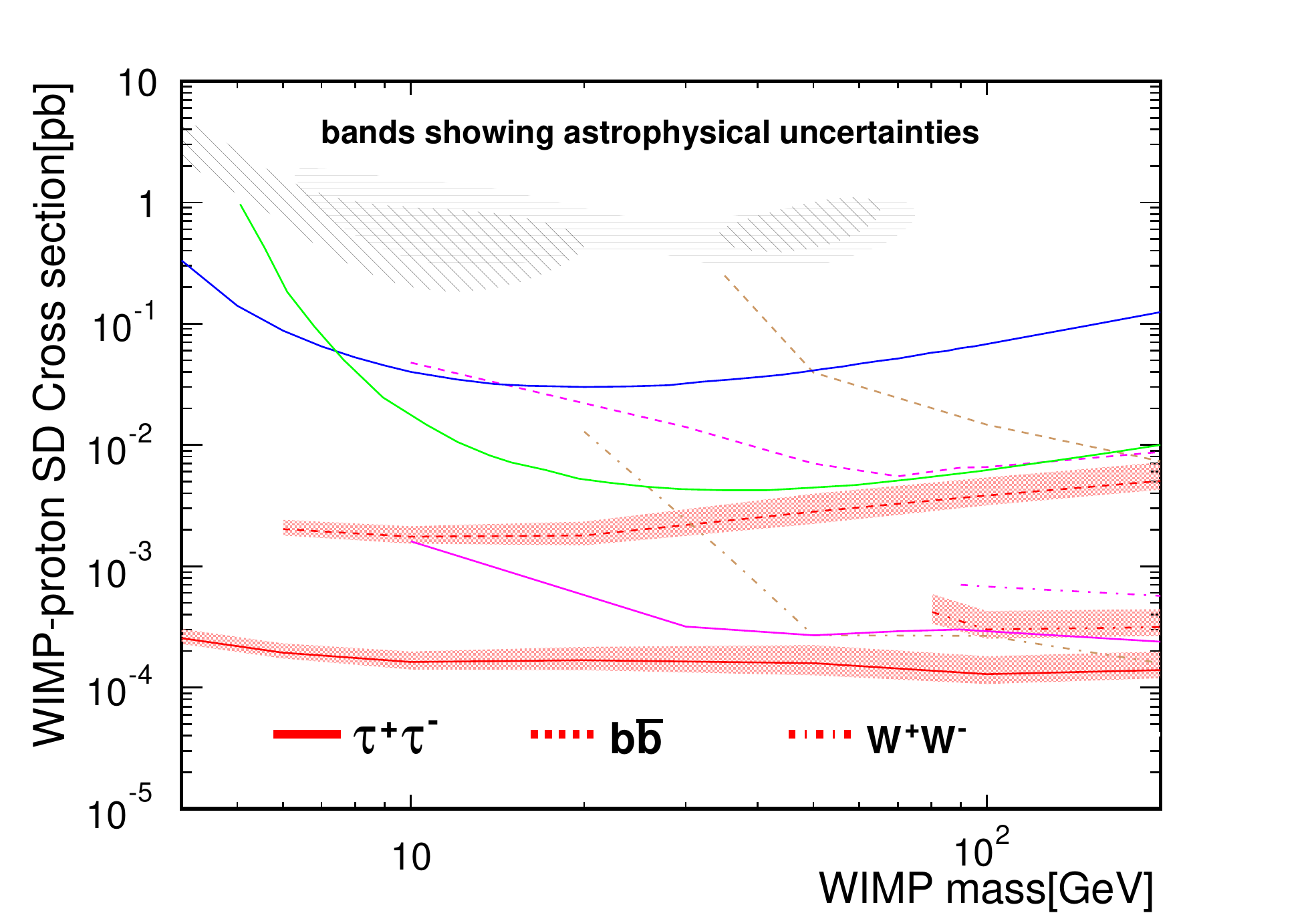}
\caption{Upper limits on the SD WIMP-proton cross section calculated using DarkSUSY~\cite{Gondolo:2004sc}. The 90\% C.L. limits for the 
         Super-K results are: $\tau^+\tau^-$ (red solid), $b\overline{b}$ (red dashed); $W^+W^-$ (red dot-dashed).
The bands around each limit indicate the uncertainty in the assumed solar model and velocity distribution. Also shown are limits from other experiments: IceCube~\cite{Aartsen:2012kia} $b\overline{b}$ (brown dashed) / $W^+W^-$ (brown dot-dashed); BAKSAN~\cite{Boliev:2013ai} $\tau^+\tau^-$ (pink solid) / $b\overline{b}$ (pink dashed) / $W^+W^-$ (pink dot-dashed); PICASSO~\cite{Archambault:2012pm} (blue solid); SIMPLE~\cite{Felizardo:2011uw} (green solid). The black shaded regions are the $3\sigma$ C.L. signal claimed by DAMA/LIBRA~\cite{Bernabei:2008yi,Savage:2008er}.}
\label{fig:solwimp}
\end{figure}

WIMP annihilation into several channels is considered. 
For annihilation in the GC these channels are  $\chi\chi \rightarrow W^{+}W^{-} , b \bar b, \mu^{+}\mu^{-}$ and $\nu \bar \nu$, 
and for annihilation in the Sun they are  $\chi\chi \rightarrow W^{+}W^{-} , b \bar b, \tau^{+}\tau^{-}$.
Due to the strong directionality of the expected signal, these analyses are binned 
in both momentum as well as the cosine of the direction to the GC (Sun).
Figure~\ref{fig:gcwimpsignal} shows an illustration of the expected signal at Super-K 
for 5 GeV WIMP annihilation overlaid with the data and MC assuming standard oscillations.
The horizontal axis shows the cosine of the direction to the GC and the signal normalization 
is arbitrary.  
Though the atmospheric neutrino data and MC distribution do not peak towards the GC, the simulated WIMP signal does; this feature 
allows for powerful constraints on a potential signal.
For annihilation in the Sun the plots look similar, though the direction coordinate is different.

The analyses are performed by assuming several potential WIMP masses ranging from a few to several hundred 
GeV and separately assuming a 100\% branching ratio into each of the annihilation channels above. 
During the fit the signal and atmospheric neutrino background normalizations are simultaneously 
varied to achieve the best agreement with the data.
No evidence for an excess of WIMP-induced neutrinos is found in either analysis.
As a result, for the GC analysis, limits are placed on the velocity averaged WIMP self-annihilation 
cross section using the DM density distribution given by~\cite{Navarro:1995iw}.
Figure~\ref{fig:gcwimp} shows the 90\% C.L. upper bounds on this parameter for a variety 
of annihilation channels and assumed dark matter masses.

By making assumptions about the composition of the Sun and the velocity distribution of 
WIMPs trapped in its gravitational potential, it is possible to extract limits on both 
the WIMP spin-dependent (SD) and spin-independent (SI) scattering cross section on 
nuclei. 
Assuming that the capture rate of WIMPs is in equilibrium with their annihilation rate, 
that they populate a DM halo with local density 0.3 GeV/cm$^3$~\cite{Kamionkowski:1997xg,Yao:2006px}
with an RMS velocity of 270 km/s, and that the solar rotation speed is of 220 km/s,
the SD limits appearing Figure~\ref{fig:solwimp} have been obtained.
Note that these are the most stringent constraint on the SD cross section 
for WIMP mass less than 200 $\mbox{GeV/c}^{2}$ assuming 100\% annihilation into
$\tau^{+}\tau^{-}$.
This analysis also places stringent limits on the SI cross section. 
For instance, the 90\% C.L. limits on $\chi\chi \rightarrow b \bar b$ and 
$\chi\chi\rightarrow \tau^{+}\tau^{-}$ are  $240 \times 10^{-43}$cm$^2$  
and $21.2 \times 10^{-43}$cm$^2$,respectively.
These results are in tension with the findings of CDMS~II~\cite{Agnese:2013rvf} 
and to a lesser extent with 
CoGeNT~\cite{Aalseth:2010vx} and CRESSTII~\cite{Angloher:2011uu}.

%10 & $b \overline{b}$ &  14.9 &  240 \\
%\cline{2-4}
%& $\tau^+\tau^-$ & 1.31 &21.2 \\
%\hline

%
%\begin{table}
%\begin{tabular}{lccc}
%\hline
%\hline
%$m_{\chi}$ & annihilation & $\sigma_{SD,p}$           & $\sigma_{SI,p}$   \\
%(GeV/$c^2$) & channel     & ($\times 10^{-40}$cm$^2$) & ($\times 10^{-43}$cm$^2$) \\
%\hline
%4 & $\tau^+\tau^-$   &  2.22     & 87.3\\
%\hline
%6 & $b \overline{b}$ &  17.2        & 456\\
%\cline{2-4}
% & $\tau^+\tau^-$    &   1.63  &44.5 \\
%\hline
%10 & $b \overline{b}$ &  14.9 &  240 \\
%\cline{2-4}
%& $\tau^+\tau^-$ & 1.31 &21.2 \\
%\hline
%20 & $b \overline{b}$ &     14.3   & 120 \\
%\cline{2-4}
%& $\tau^+\tau^-$ &   1.42   &11.9 \\
%\hline
%50 & $b \overline{b}$ &  23.4   & 89.9 \\
%\cline{2-4}
%& $\tau^+\tau^-$ &      1.28   & 4.92 \\
%\hline
%80.3 & $W^+W^-$ &  3.13 & 8.26 \\
%\hline
%100 & $b \overline{b}$ & 31.9 & 71.3 \\
%\cline{2-4}
%& $W^+W^-$ &   2.80  & 6.26 \\
%\cline{2-4}
%& $\tau^+\tau^-$ & 1.24 & 2.76 \\
%\hline
%200 & $b \overline{b}$ &          44.9  & 63.4 \\
%\cline{2-4}
%& $W^+W^-$ &    3.00 &      4.23 \\
%\cline{2-4}
%& $\tau^+\tau^-$  & 1.33 & 1.88\\
%\hline
%\hline
%\end{tabular}
%\caption{$\chi^2_{min}$, $\Delta\chi^2$ at $\beta=0$, $\Delta\chi^2_{90}$ ($\Delta\chi^2$ for 90\% Bayesian upper limit), 90\% upper limit on the muon-neutrino flux from WIMP annihilations in the Sun at SK and SD/SI/SI~(IVDM) scattering cross section limits for each WIMP mass and annihilation channel.
%}
%\label{table1}
%\end{table}

\subsubsection{Conclusion}
Super-K atmospheric neutrino data have been studied in the context of 
both standard and exotic oscillation scenarios as well as in the search 
for neutrinos produced in the annihilation of WIMPs in our galactic halo 
and Sun. 
Atmospheric neutrino constraints on standard mixing remain compelling and 
thus far have a weak, $\sim$1$\sigma$ preference for the normal mass hierarchy. 
While this result remains statistically limited, efforts to improve sensitivity 
by introducing constraints from external experiments have yielded similar conclusions. 
Though no evidence for the existence of sterile or LV oscillations is found 
in the data, stringent limits on the parameters governing both have been 
obtained.
Finally, searches for an excess of neutrinos coming from the direction of the 
galactic center and the Sun, have resulted in tight constraints, in particular 
at low masses, on 
the WIMP's velocity averaged self-annhilation cross section as well as its 
scattering cross sections on nuclei.

\begin{theacknowledgments}
The author would like to thank the organizers for the opportunity to present these results.
The Super-Kamiokande collaboration gratefully acknowledges the cooperation of the Kamioka
Mining and Smelting Company. Super-K has been built and operated from
funds provided by the Japanese Ministry of Education, Culture, Sports,
Science and Technology, the U.S.  Department of Energy, and the
U.S. National Science Foundation. This work was partially supported by
the Research Foundation of Korea (BK21 and KNRC), the Korean Ministry
of Science and Technology, the National Science Foundation of China,
the European Union FP7 (DS laguna-lbno PN-284518 and ITN
invisibles GA-2011-289442)
the National Science and Engineering Research Council (NSERC) of Canada,
and the Scinet and Westgrid consortia of Compute Canada.
\end{theacknowledgments}

\bibliographystyle{aipproc}   % if natbib is available
%\bibliographystyle{aipprocl} % if natbib is missing

%%%%%%%%%%%%%%%%%%%%%%%%%%%%%%%%%%%%%%%%%%%
%% You probably want to use your own bibtex database here
%%%%%%%%%%%%%%%%%%%%%%%%%%%%%%%%%%%%%%%%%%%
\bibliography{wendell_sk_nu2014}

%%%%%%%%%%%%%%%%%%%%%%%%%%%%%%%%%%%%%%%%%%%
%% Just a reminder that you may have to run bibtex
%% All of it up to \end{document} can be removed
%% if you don't like the warning.
%%%%%%%%%%%%%%%%%%%%%%%%%%%%%%%%%%%%%%%%%%%
\IfFileExists{\jobname.bbl}{}
 {\typeout{}
  \typeout{******************************************}
  \typeout{** Please run "bibtex \jobname" to optain}
  \typeout{** the bibliography and then re-run LaTeX}
  \typeout{** twice to fix the references!}
  \typeout{******************************************}
  \typeout{}
 }

\end{document}